\documentclass[3p,times,twocolumn]{elsarticle}

\usepackage{ecrc}

\usepackage{graphicx}  

\usepackage{amsmath}
\usepackage{multirow}
\usepackage{setspace}
\usepackage{hhline}

\def\Journal#1#2#3#4{{#1} {\bf #2}, #3 (#4)}

\def\PRL{\em Phys. Rev. Lett.}
\def\PRD{{\em Phys. Rev.} D}

\volume{00}

\firstpage{1}

\journalname{Nuclear Physics B Proceedings Supplement}

\runauth{Chang-Seong Moon}

\jid{nuphbp}

\jnltitlelogo{Nuclear Physics B Proceedings Supplement}

\usepackage{amssymb}

\biboptions{compress}

\usepackage[figuresright]{rotating}

\begin{document}

\begin{frontmatter}



\dochead{}

\title{Top quark pair production and top quark properties at CDF}


\author{Chang-Seong Moon, on behalf of the CDF Collaboration}

\address{AstroParticule et Cosmologie, Universit\'e Paris Diderot, CNRS/IN2P3,\\
10, rue Alice Domon et L\'eonie Duquet, 75205 Paris, France\\
and INFN Sezione di Pisa, Largo B. Pontecorvo, 3, 56127 Pisa, Italy}

\begin{abstract}

We present the most recent measurements of top quark pairs production and top quark properties 
in proton-antiproton collisions with center-of-mass energy of 1.96 TeV using CDF II detector at 
the Tevatron.
The combination of top pair production cross section measurements and 
the direct measurement of top quark width are reported. 
The test of Standard Model predictions for top quark decaying into $b$-quarks, 
performed by measuring the ratio $R$ between the top quark branching fraction to $b$-quark and 
the branching fraction to any type of down quark is shown.
The extraction of the CKM matrix element $|V_{tb}|$ from the ratio $R$ is discussed. 
We also present the latest measurements on the forward-backward asymmetry ($A_{FB}$) in top anti-top 
quark production. With the full CDF Run II data set, the measurements are performed 
in top anti-top decaying to final states that contain one or two charged leptons (electrons or muons). 
In addition, we combine the results of the leptonic forward-backward asymmetry in $t\bar t$ system 
between the two final states. 
All the results show deviations from the next-to-leading order (NLO) standard model (SM) calculation.

\end{abstract}


\end{frontmatter}


\section{Introduction}
\label{Introduction}

The top quark, discovered at the Tevatron in 1995 by CDF~\cite{topCDF} and D0~\cite{topD0} experiments, 
have completed the third generation of quarks in the SM and been playing 
a special role with the heaviest known elementary particle in electroweak symmetry breaking.
This large mass is responsible for the top quark properties. 
Any deviation from the SM predictions of the top quark properties could indicate the presence of 
beyond SM signal indirectly.

Top quarks are mostly produced in top-antitop pairs via the strong interaction at the Tevatron.
Thousands of top events have been collected by CDF Run II detector. It allows performing precision measurements 
of top properties and many analyses are unique to the Tevatron and complementary to the LHC measurements.

In the SM, the top quark decays to a $W$-boson and a $b$-quark almost 100 \% of the time.
The decay channels of the top quark are labeled as dilepton, lepton+jets and all hadronic channel classified 
by whether a leptonical decay into electron or muon has occurred in both (5 \%), one only (30 \%) or none of 
the two $W$-bosons (45 \%) respectively.
The dilepton channel is only 5 percent of top quark pairs decay but it has the cleanest final state
to identify because there is no so many other processes can produce such a striking final state 
even before requiring the identification of one of the jets originating from a $b$-quark ($b$-tagging).
The lepton+jets mode is referred to as the golden channel because of a sizable branching fraction 
and controllable background levels using the $b$-tagging. The all hadronic decay channel 
has an advantage of the largest cross section but it suffers from a huge background from the
quantum chromodynamics (QCD) multijet production and cannot be easily simulated with Monte Carlo sample.

\section{Top quark width and branching fraction ratio}
  \label{Top quark width and branching fraction ratio}

\begin{figure}
\begin{minipage}{0.99\linewidth}
\centerline{\includegraphics[width=0.99\linewidth]{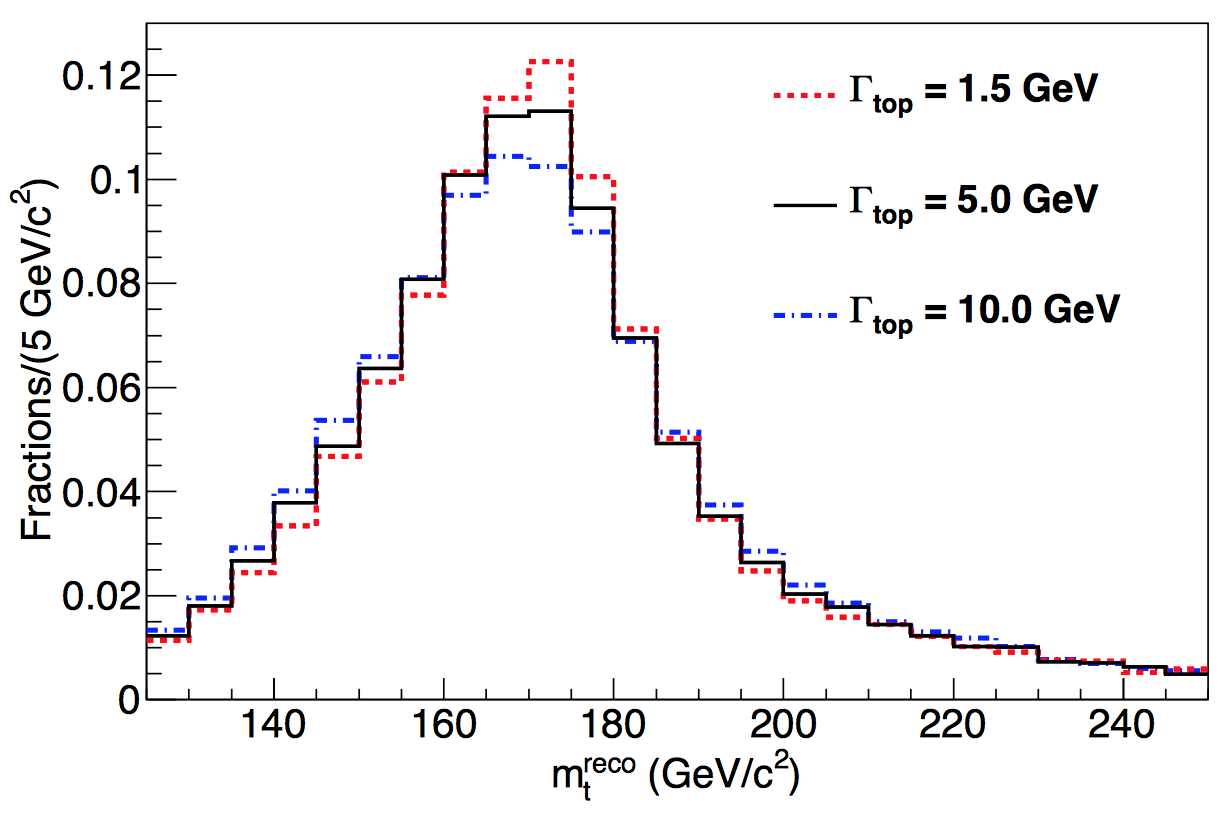}}
\vspace{3mm}
\centerline{\includegraphics[width=0.99\linewidth]{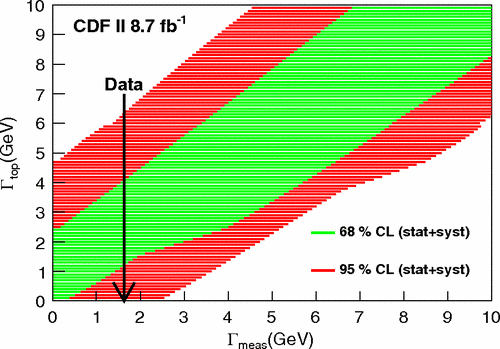}}
\vspace{3mm}
\centerline{\includegraphics[width=0.99\linewidth]{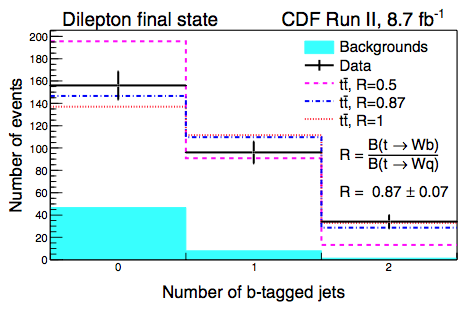}}
\end{minipage}
\caption[]{
(top) 
$m^{reco}_t$ distributions for simulated events meeting the lepton + jets selection
are displayed with three values of top and with the nominal jet-energy scale.
(middle) 
Confidence bands of $\Gamma_{top}$ as a function of $\Gamma_{meas}$ for 68 \% and 95 \% CL limits. 
Results from simulated experiments assuming 8.7 fb$^{-1}$ of data at different values of 
top are convoluted with a smearing function to account for systematic uncertainties. 
The value observed in data is indicated by an arrow.
(bottom) 
Number of events observed in data and expected for various values of $R$ as a function of 
identified $b$-jets.
}
\label{fig:top_width}
\end{figure}

\setstretch{0.97}

Top quark has the largest decay width of the known fermions in the SM. 
The recent next-to-next-to-leading-order (NNLO) calculation with QCD and electroweak corrections predicts 
$\Gamma_{top}$, the top quark decay width to be 1.32 GeV at top quark mass of $M_{top}$ = 172.5 GeV/c$^2$~\cite{topWidth}. 
Direct measurements of the top quark width are performed in fully reconstructed lepton+jets events 
using the full CDF Run II data set, corresponding to an integrated luminosity of 8.7 fb$^{-1}$.
The best-fit value of measured top quark width ($\Gamma_{meas}$) retrieved from the data is 1.63 GeV 
and is depicted as an arrow in the plot of Fig.~\ref{fig:top_width} (middle), 
This corresponds to an upper limit of $\Gamma_{top} <$ 6.38 GeV at the 95 \% confidence level (CL)
We also set a two-sided limit of 1.10 GeV $< \Gamma_{top} < $4.05 GeV at 68 \% CL,
which corresponds to a lifetime of 1.6$\times$10$^{-25}$ s, this result confirms the prediction that top quark 
decays before its hadronization.
This is the most precise direct determination of the top quark width and lifetime and 
shows no evidence of non-SM physics in the top quark decay~\cite{cdf_top_width}.

In the SM, the top quark coupling to a down-type quark $q$, where $q$ = $d$, $s$, $b$ is 
proportional to $|V_{tq}|$, the element of the Cabibbo-Kobayashi-Maskawa (CKM) matrix. 
Under the assumption that the 3$\times$3 CKM matrix is unitary and given constraints on $|V_{ts}|$ and $|V_{td}|$,
the magnitude of the top-bottom quark coupling is computed as $|V_{tb}|$ = 0.99915 $^{+0.00002} _{-0.00005}$~\cite{BR}.
Using these hypotheses, the ratio of the top quark branching fraction $R$ is indirectly determined as
%
%
\begin{equation}
\begin{split}
R = \frac{B(t\rightarrow Wb)}{B(t\rightarrow Wq)} 
&= \frac{|V_{tb}|^2}{|V_{tb}|^2+|V_{tb}|^2+|V_{tb}|^2} \\ 
&= 0.99830 ^{+0.00004} _{-0.00009}
\end{split}
\label{eq:R}
\end{equation}
%
%
The ratio of the top quark branching fraction $R$ is recently measured 
in events with two charged leptons, imbalance in total transverse energy, 
and at least two jets (dilepton channel).
The sample is divided into 9 subsamples by dilepton flavor ($ee$, $\mu\mu$, $e\mu$) 
and 3 $b$-tagging categories (presence of 0, 1, or 2 tags)
in order to better exploit the subsample-dependent signal-to-background ratio and 
compare between observed data and expectations.
The number of observed and predicted events are used in the various subsamples as input to a likelihood function, 
which is maximized to extract $R$, since the number of $b$-jets in the event is related to the top quark branching 
fraction in the $W b$ final state. 
We obtain $R >$ 0.73(0.76) at 95 \% (90 \%) CL. From Eq.~(\ref{eq:R}) and the assumptions therein we obtain 
$|V_{tb}|$ = 0.94 $\pm$ 0.04 and $|V_{tb}| >$ 0.85(0.87) at 95 \% (90 \%) CL.
The result, $R$ = 0.87 $\pm$ 0.07 differs from the SM expectation by $\approx$ 1.8$\sigma$~\cite{dilBR}.

\setstretch{1.00}

\section{Top quark pair production cross section}
  \label{Top quark pair production cross section}

\begin{figure*}[htbp]
 \begin{center}
  \begin{tabular}[t]{cc}
    \includegraphics[width=0.47\textwidth,height=0.40\textwidth, clip]{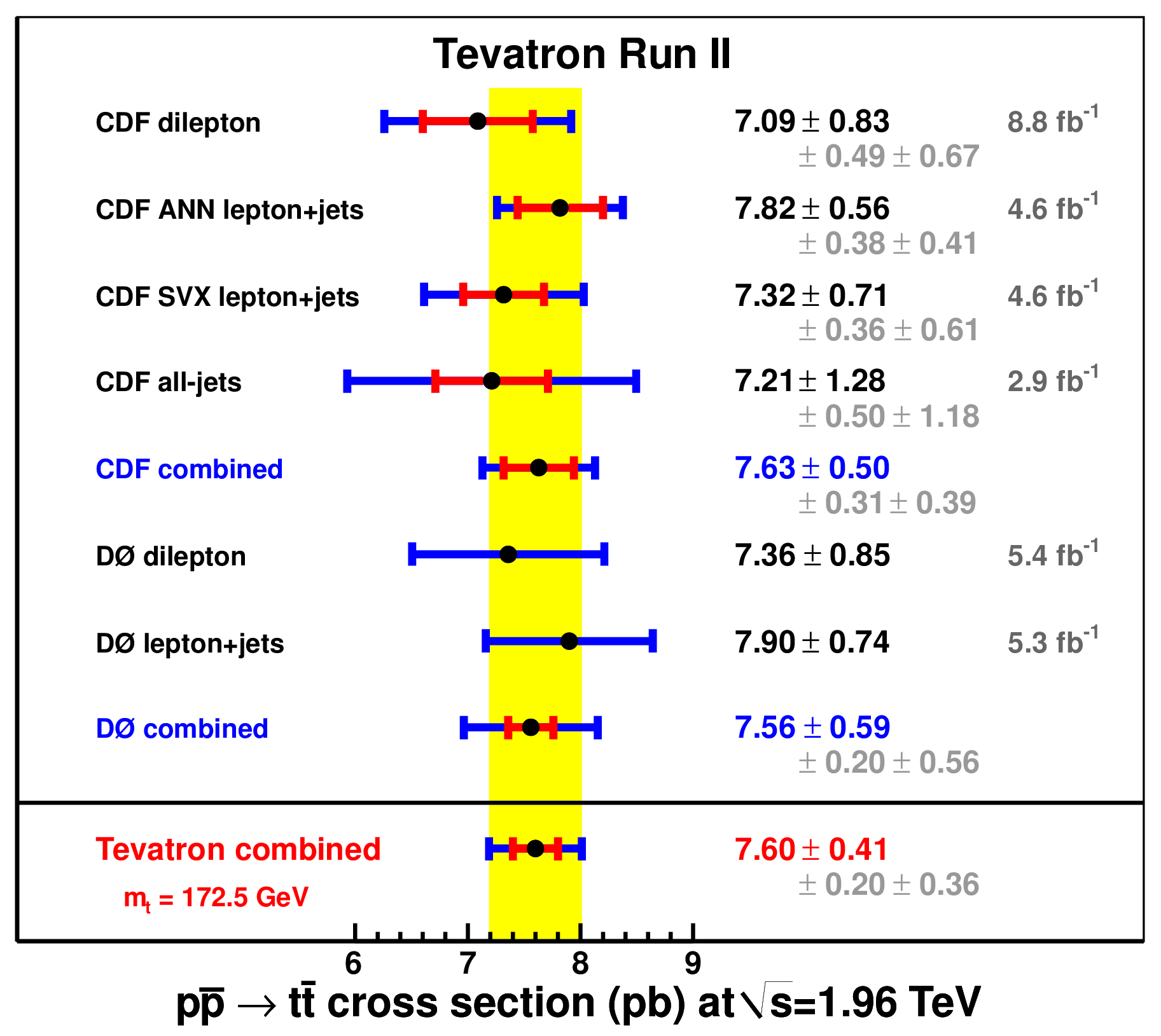} &
    \includegraphics[width=0.47\textwidth,height=0.395\textwidth, clip]{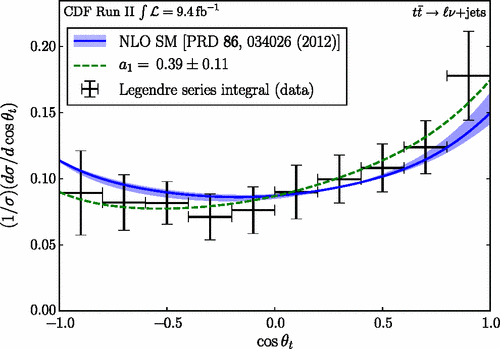} \\
  \end{tabular}
 \end{center}
  \caption{ 
(left) 
The CDF-only combination result with the four input $\sigma_{t\bar t}$ measurements from the CDF experiment, 
along with the D0-only combination for the combination for the Tevatron result. 
The inner (red) bars reflect statistical uncertainties while the 
outer (blue) bars show the total uncertainties on each measurement.
(right)
Fraction of cross section in 10 bins of cos$\theta_t$, 
obtained by integrating the series of Legendre polynomials over the width of each bin.
}
\label{fig:xsec}
\end{figure*}

CDF collaboration recently combined four measurements of the inclusive top quark pair ($t\bar t$) production 
cross section ($\sigma_{t\bar t}$) from the full CDF Run II data set corresponding to integrated luminosities 
of up to 8.8 fb$^{-1}$ by contribution of a measurement in the dilepton channel~\cite{dilXS}. 
We obtain a value of $\sigma_{t\bar t}$ = 7.63 $\pm$ 0.50 pb for a top quark mass of $M_{top}$ = 172.5 GeV. 
The contributions to the uncertainty are 0.31 pb from statistical sources, 0.39 pb from systematic sources 
including the uncertainty on the integrated luminosity. 
The result is in good agreement with the standard model expectation of 7.35 $^{+0.28} _{-0.33}$ pb
at NNLO and next-to-next-to leading logarithms in perturbative QCD~\cite{SMXS}.
The CDF-only combination result contributes to 60 \% of the combination for the Tevatron result to be
$\sigma_{t\bar t}$ = 7.60 $\pm$ 0.41 pb~\cite{XS}.

A measurement of the differential $t\bar t$ cross section, $d\sigma/d(\text{cos}\theta_t)$ for top quark pair 
production as a function of the top quark production angle in proton-antiproton collisions is reported~\cite{dXS}.
This measurement is performed using the lepton+jets events and the full CDF Run II data set corresponding to 
an integrated luminosity of 9.4 fb$^{-1}$. 
We employ the Legendre polynomials to characterize the shape of the differential cross section at the parton-level. 
We characterize the shape of $d\sigma/d(\text{cos}\theta_t)$ by expanding in the Legendre polynomials
where $P_{\ell}$ is the Legendre polynomial of degree $\ell$, and $a_{\ell}$ is the Legendre moment of 
degree $\ell$ in Eq.~(\ref{eq:dsigma}).
\begin{equation}
\frac{d\sigma}{d(\text{cos}\theta_t)} = \sum^{\infty}_{\ell=0}a_{\ell}P_{\ell}(\text{cos}\theta_t)
\label{eq:dsigma}
\end{equation}
The observed Legendre coefficient of the cos$\theta_t$ linear-term in the differential cross section
$a_1$ = 0.40 $\pm$ 0.12 is excess of the NLO SM prediction of $a_1$ = 0.15 $^{+0.07} _{-0:03}$,
while the remainder of the differential cross section is in good agreement within the uncertainties 
with the prediction of the NLO SM calculation~\cite{dXS}.

\section{Top quark forward-backward asymmetry}
  \label{Top quark forward-backward asymmetry}

The forward-backward asymmetry ($A_{FB}$) in $p\bar p\rightarrow t\bar t$ production is defined as
\begin{equation}
A_{FB} = \frac{N(\Delta y>0)-N(\Delta y<0)}{N(\Delta y>0)+N(\Delta y<0)}~(\Delta y = y_t - y_{\bar{t}}),
\end{equation}
where $y$ is given by $y=\frac{1}{2}ln\left(\frac{E+p_z}{E-pz}\right)$
with $E$ being the total top quark energy and $p_z$ being the component of the top quark momentum 
along the beam axis as measured in the detector rest frame. 
The $A_{FB}$ is identical to the asymmetry in the top-quark production angle in the experimentally 
well-defined $t\bar t$ rest frame.
An NLO SM calculation with both electroweak and NLO QCD effects lead to a small $A_{FB}$ of order 
8.8 $\pm$ 0.6 \% in the $t\bar t$ rest frame~\cite{afbSM}.
Asymmetries in $t\bar t$ production beyond SM prediction could provide 
the first evidence of new interactions, such as $t\bar t$ production via a heavy axial color octet or 
a flavor-changing Z$^\prime$ boson that might not be easily observed as excesses in the top-quark production rate 
or as resonances in the $t\bar t$ invariant mass distribution~\cite{afb_lj}.

CDF measured a reconstruction-level asymmetry of $A_{FB}$ = 6.3 $\pm$ 1.9 \% and 
an inclusive parton-level $A_{FB}$ = 16.4 $\pm$ 4.7 \% in the lepton+jets decay channel
with the full CDF data set corresponding to an integrated luminosity of 9.4 fb$^{-1}$.
In addition, differential measurements using two bins each in the top-antitop rapidity difference $|\Delta y|$ 
and the top-antitop invariant mass $M_{t\bar t}$ are performed. 
The results agree on a larger dependence of $A_{FB}$ on $|\Delta y|$ and $M_{t\bar t}$ than the NLO SM expectation.
The observed slopes of the asymmetries by linear fit of the data as a function of $|\Delta y|$ and $M_{t\bar t}$ 
give 2.8 $\sigma$ and 2.4 $\sigma$ excesses from the theoretical prediction respectively. 
We also study the dependence of the asymmetry on the $t\bar t$ transverse momentum of the system 
$p^{t\bar t}_T$ at the detector level. 
These results show the excess asymmetry in the data is consistent with being dependent of $p^{t\bar t}_T$ and 
provide additional quantification of the functional dependencies of the asymmetry~\cite{afb_lj}.

\section{Leptonic asymmetry in $t\bar t$ production}
  \label{Leptonic asymmetry in ttbar production}

The leptonic forward-backward asymmetry $A^{\ell}_{FB}$ ($\ell$ = $e$ or $\mu$)
kinematically correlated with the top quark pair asymmetry $A^{t\bar t}_{FB}$ as well as the top quark polarizations.
If the genuine value of $A^{t\bar t}_{FB}$ would be that measured by CDF collaboration~\cite{afb_lj}, 
the predicted value for $A^{\ell}_{FB}$ for top quarks decaying according to the SM would be 
7.0 \% $< A^{\ell}_{FB}$ $<$ 7.6 \%~\cite{lepAfb_lj}.
The inclusive asymmetry of the charge-weighted lepton pseudorapidities ($q_{\ell}y_{\ell}$)
is defined as 
\begin{equation}
A^{\ell}_{FB} = \frac{N(q_{\ell}y_{\ell}>0)-N(q_{\ell}y_{\ell}<0)}{N(q_{\ell}y_{\ell}>0)+N(q_{\ell}y_{\ell}<0)}
\label{eq:lepasym}
\end{equation}
The asymmetric part is decomposed from $q_{\ell}y_{\ell}$ distribution at parton-level 
with various physics models~\cite{lepAfb_lj,lepAfb_dil}. 
The differential asymmetry as a function of $q_{\ell}y_{\ell}$, $\mathcal{A}$($q_{\ell}y_{\ell}$) 
is parametrized with an functional form of 
\begin{equation}
\mathcal{A}(q_{\ell}y_{\ell}) = a\cdot\mathrm{tanh} \left(\frac{q_{\ell}y_{\ell}}{2}\right),
\label{eq:dasym}
\end{equation}
where $a$ is the only free parameter related to $A^{\ell}_{FB}$.
A predicted value in the SM is $A^{\ell}_{FB}$ = 3.8 $\pm$ 0.3 \%~\cite{afbSM}. 

A second observable for the leptonic asymmetry is defined in the dilepton final state analogously to $A^{t\bar t}_{FB}$ as 
\begin{equation}
A^{\ell\ell}_{FB} = \frac{N(\Delta y>0)-N(\Delta y<0)}{N(\Delta y>0)+N(\Delta y<0)}
\label{eq:dasym}
\end{equation}
where $\Delta y$ = $y_{\ell^+}$ - $y_{\ell^-}$. 
An NLO SM prediction yields $A^{\ell\ell}_{FB}$ = 4.8$\pm$ 0.4~\cite{afbSM}.

\begin{figure}[htbp]
\centering
\begin{minipage}{0.99\linewidth}
\centerline{\includegraphics[width=0.99\linewidth]{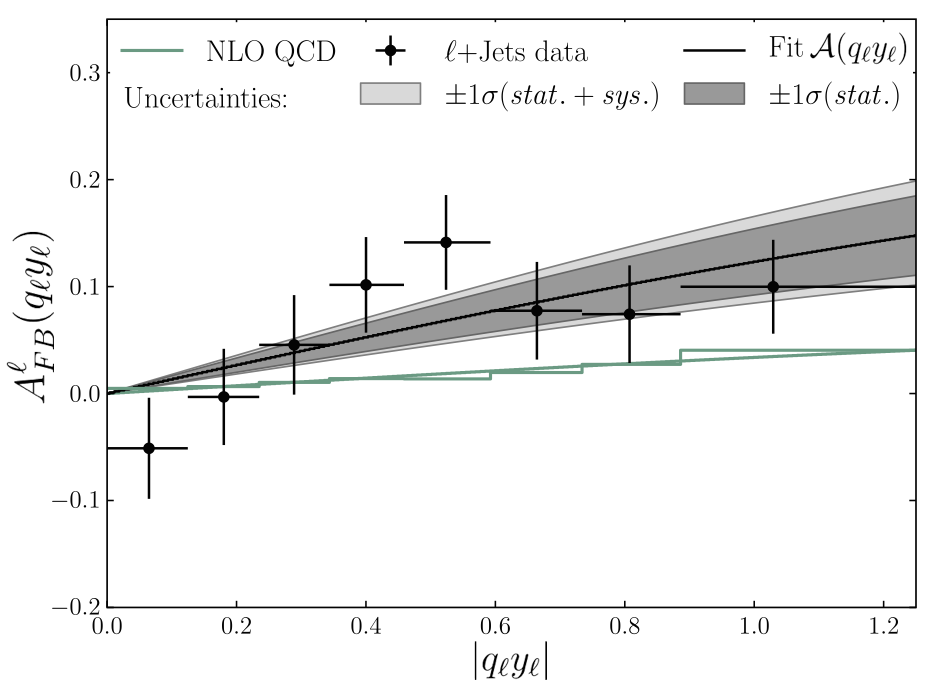}}
\vspace{3mm}
\centerline{\includegraphics[width=0.97\linewidth]{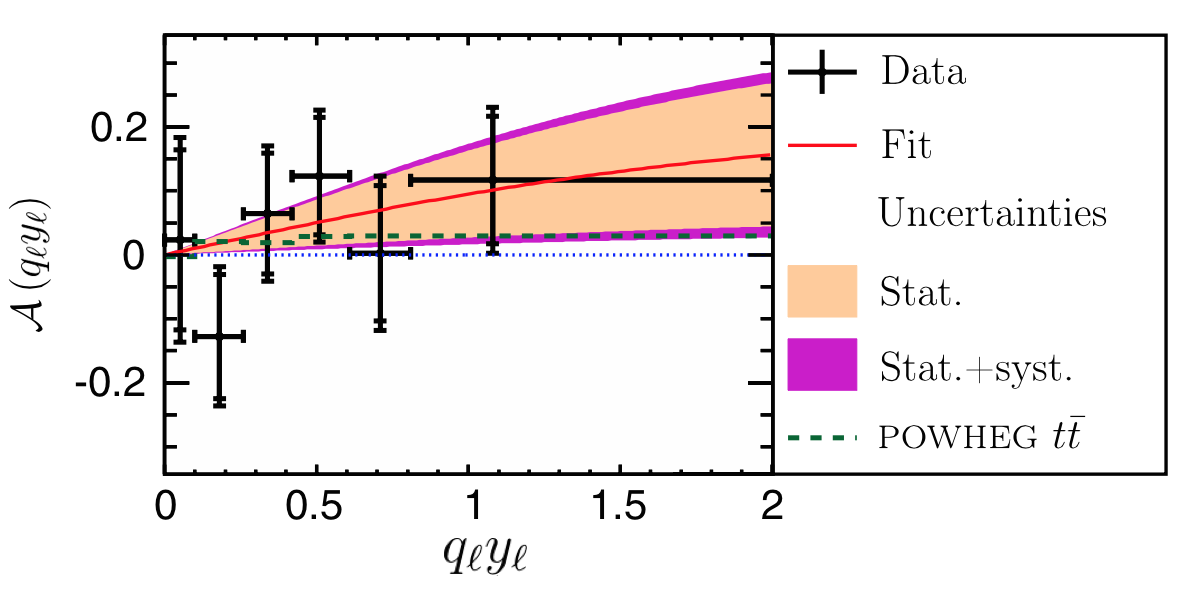}}
\end{minipage}
\caption[]{
(top) 
The binned asymmetry $A^{\ell}_{FB}(q_{\ell}y_{\ell})$ in a single charged lepton and hadronic jets (lepton+jets)
after correcting for acceptance, 
compared to the NLO QCD prediction of POWHEG. 
The best fit to Eq. (\ref{eq:dasym}) for each is shown as the smooth curve of the same color. 
The dark (light) gray bands indicate the statistical (total) uncertainty on the fit curve to the data.
The data points for the final state with two charged leptons (dilepton) 
in (bottom) are placed at the bin centroids predicted by the POWHEG simulation. 
The inner bars on the data points represent the statistical uncertainties, 
while the outer bars represent the total uncertainties. 
The bands indicate the one standard deviation region for statistical and statistical + systematic uncertainties.
}
\label{fig:leptonic_afb}
\end{figure}

The first measurement of the production-level inclusive lepton asymmetry in the lepton+jets events 
is found to be $A^{\ell}_{FB}$ = 9.4$^{+3.2}_{-2.9}$ \% using the full CDF data set is reported~\cite{lepAfb_lj}
as shown in Fig.~\ref{fig:leptonic_afb} (top).
Following measurements of the parton-level inclusive leptonic forward-backward asymmetry 
of top-quark pairs decaying into the dilepton final state using the full CDF data set 
are presented~\cite{lepAfb_dil}.
The results are $A^{\ell}_{FB}$ = 7.2 $\pm$ 6.0 \% and $A^{\ell\ell}_{FB}$ = 7.6 $\pm$ 8.2 \%
as described in Fig.~\ref{fig:leptonic_afb} (bottom), 
both consistent with the previous determination~\cite{lepAfb_lj} and the SM expectations~\cite{afbSM}.

\begin{figure}[htbp]
\centering
\begin{minipage}{0.99\linewidth}
\centerline{\includegraphics[width=1.03\linewidth]{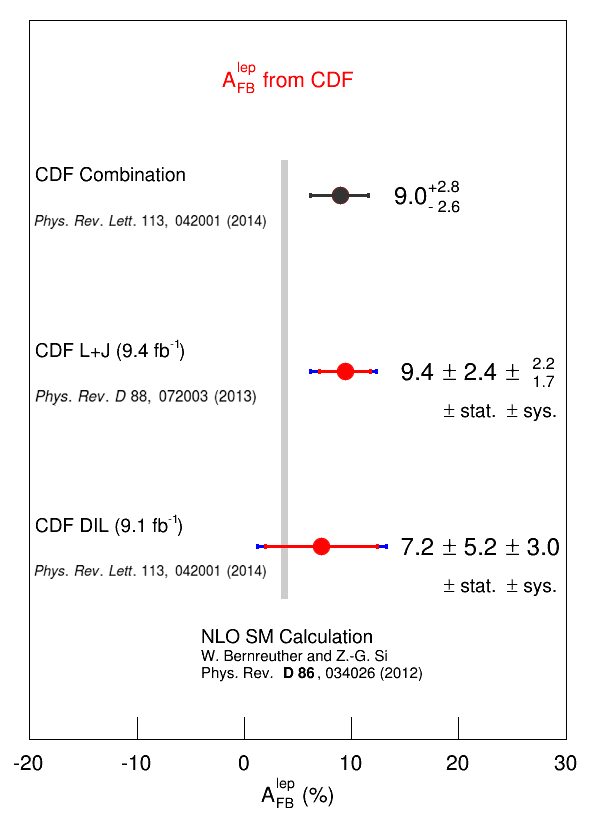}}
\end{minipage}
\caption[]{
The measurements of $A^{lep}_{FB}$ with leption+jets and dilepton final states from CDF, 
along with their combination for the CDF result.
}
\label{fig:leptonic_afb2}
\end{figure}

A combination of CDF $A^{\ell}_{FB}$ measurements with the dilepton and lepton+jets events yields 
$A^{\ell}_{FB}$ = 9.0$^{+2.8}_{-2.6}$,
where 80 \% of the measurement weight is due to the lepton + jets result and 20 \% is due to the dilepton result. 
The difference in the weights is mostly due to the larger size of the lepton + jets final state sample. 
This result is about two standard deviations larger than the NLO SM calculation of 
$A^{\ell}_{FB}$ = 3.8 $\pm$ 0.3 \%~\cite{lepAfb_dil},
but is consistent with the 7.0-7.6 \% range expected assuming unpolarized top quark production
and SM top-quark decay, given the measured value of $A^{\ell}_{FB}$ by the CDF collaboration~\cite{lepAfb_lj}.
These CDF results of $A^{\ell}_{FB}$ measurements are compared in Fig.~\ref{fig:leptonic_afb2}.

\section{Forward-backward asymmetry in $b\bar b$ pairs}
  \label{Forward-backward asymmetry in bbbar pairs}

\begin{figure}[htbp]
\begin{minipage}{0.99\linewidth}
\centerline{\includegraphics[width=0.99\linewidth]{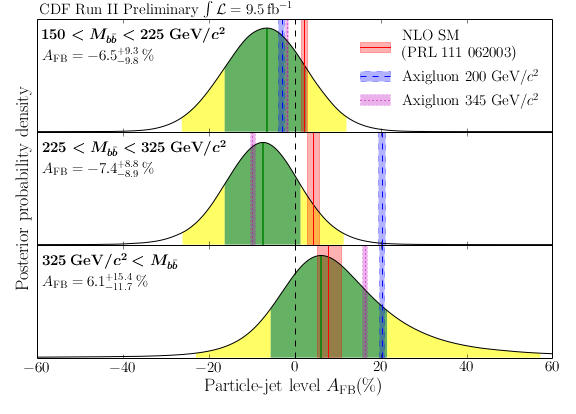}}
\vspace{3mm}
\centerline{\includegraphics[width=0.97\linewidth]{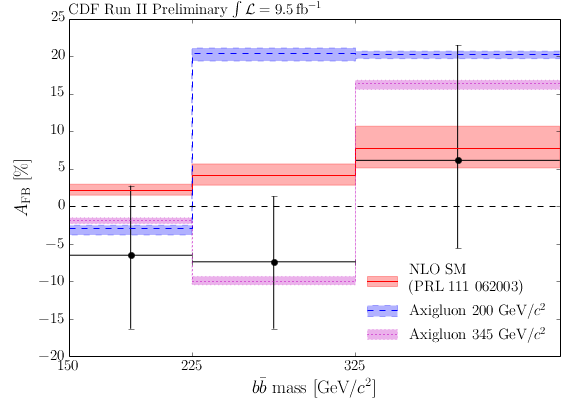}}
\end{minipage}
\caption[]{
(top) 
Marginal posterior probability distribution of asymmetry in each bin of particle-level $b\bar b$ mass. 
The green and yellow bands represent the 68 \% and 95 \% credible intervals, respectively.
(bottom)
Maximum a posteriori points for the signal asymmetry in each mass bin. The error bars represent 
the 68 \% credible intervals.
}
\label{fig:afb_bbar}
\end{figure}

The light axigluon models which are viable candidates to explain the forward-backward asymmetry in $t\bar t$ production, 
predict forward-backward asymmetries for $b\bar b$ and $c\bar c$ systems~\cite{afb_bbar1,afb_bbar2,afb_bbar3,afb_bbar4}. 
The $A_{FB}$ in $b\bar b$ pairs at large $b\bar b$ mass ($M_{b\bar b}>$150 GeV) is first measured
using jet-triggered data and jet charge to identify $b$-quark from $\bar b$-quark.
To characterize the posterior and describe the measured data, 
we find the highest-probability-density credible intervals at 68 \% and 95 \% credibility 
for the $b\bar b$ asymmetry in bins of particle-level mass as displayed in Fig.~\ref{fig:afb_bbar} (top).
The observed asymmetry is consistent with both zero and with the SM prediction as a function of $M_{b\bar b}$. 
The Axigluon model which has its mass $M_A$ = 200 GeV/c$^2$ is excluded, 
while the heavier axigluon at $M_A$=345 GeV c$^2$ is good agreement with the data as shown in Fig.~\ref{fig:afb_bbar} 
(bottom)~\cite{afbBB}.
CDF is currently working on the measurement of $A_{FB}$ in $b\bar b$ pairs at low mass as well to confirm the results.

\newpage

\section{Conclusion}
  \label{Conclusion}

The recent measurements on the top quark pairs production and properties from CDF are presented. 
Most results of the top quark properties at CDF agree with the SM prediction,
while the top quark forward-backward asymmetry and the leptonic asymmetry in $t\bar t$ system 
show deviations from SM predictions.
The various different experimental checks are performed for the top quark $A_{FB}$ 
which is disagreement with the NLO SM calculation with both electroweak and NLO QCD effects.
We found the top quark asymmetry grows faster than the expectation in the NLO calculation
as a function of $t\bar t$ mass and difference of top-antitop rapidities.
CDF has been preparing to combine these $A_{FB}$ results with D0 measurements for the Tevatron combination 
of the $A_{FB}$ measurement in $t\bar t$ production.
The first measurement of the bottom quark forward-backward asymmetry at large $b\bar b$ mass is reported.
The result shows the consistency with both zero and with the SM predictions and 
slightly reduces the allowed parameter space for light axigluon models to explain the top quark forward-backward
asymmetry.

\section{Acknowledgments}
  \label{Acknowledgments}

I would like to thank my colleagues of the CDF collaboration for their continuous valuable efforts and
dedication to provide these important physics results. I also thank the organizers of ICHEP 2014
in Valencia for a productive and successful conference.

\nocite{*}
\bibliographystyle{elsarticle-num}
\bibliography{martin}

\end{document}